\documentclass{ws-rv975x65}
\usepackage{ws-rv-van}            
\makeindex
\begin{document}
\chapter[Half-Skyrmion theory for 
high-temperature superconductivity]{Half-Skyrmion theory for
high-temperature superconductivity}
\author[T. Morinari]{Takao Morinari}
\address{
Yukawa Institute for Theoretical Physics, Kyoto
University, Kyoto 606-8502, Japan}
\begin{abstract}
We review the half-Skyrmion theory for copper-oxide
high-temperature superconductivity.
In the theory, doped holes create a half-Skyrmion spin texture
which is characterized by a topological charge.
The formation of the half-Skyrmion is described
in the single hole doped system,
and then the half-Skyrmion excitation spectrum is compared with 
the angle-resolved photoemission spectroscopy results
in the undoped system.
Multi-half-Skyrmion configurations are studied 
by numerical simulations.
We show that half-Skyrmions carry non-vanishing
topological charge density below a critical hole doping concentration 
$\sim 30\%$ even in the absence of antiferromagnetic
long-range order.
The magnetic structure factor exhibits incommensurate peaks in stripe
ordered configurations of half-Skyrmions and anti-half-Skyrmions.
The interaction mediated by half-Skyrmions 
leads to $d_{x^2-y^2}$-wave superconductivity.
We also describe 
pseudogap behavior arising from the excitation
spectrum of a composite particle of 
a half-Skyrmion and doped hole.
\end{abstract}

\newcommand{\be}{\begin{equation}}
\newcommand{\ee}{\end{equation}}
\newcommand{\bea}{\begin{eqnarray}}
\newcommand{\eea}{\end{eqnarray}}

\body
\section{Introduction}
One of the most challenging problems in condensed matter physics
is to unveil the mechanism of high-temperature superconductivity
in the copper oxides.
Although it has past more than two decades 
since its discovery\cite{Bednorz1986}, no established theory exists.
The most difficult aspect is to cope with strong electron correlations:
The undoped system of high-temperature superconductors is an insulator.
Contrary to conventional band insulators,
strong Coulomb repulsion makes the system insulating.
High-temperature superconductivity occurs by doping holes 
in such a Mott insulator.\cite{ImadaFujimoriTokura1998}
The pairing symmetry is not conventional $s$-wave
but $d_{x^2-y^2}$-wave.\cite{Harlingen1995}
It is believed that electron-phonon couplings do not play 
an essential role in the mechanism of high-temperature superconductivity
because of the strong Coulomb repulsion.
Searching for a mechanism based on 
the strong electron correlation is necessary.
In this chapter, as a candidate providing such a mechanism
the half-Skyrmion theory is reviewed.

The plan of the review is as follows.
In \sref{sec_high_Tc}, we review the structure,
electronic properties, and the phase diagram of 
high-temperature superconductors.
Then, we describe the half-Skyrmion spin texture in a single
hole doped system in \sref{sec_single_hole}.
The half-Skyrmion excitation spectrum is compared
with the angle-resolved photoemission spectroscopy results
in the undoped system.
Topological character and magnetic properties of 
multi-half-Skyrmion configurations
are described in \sref{sec_multi_hs_config}.
In \sref{sec_mechanism}, we describe a mechanism of $d_{x^2-y^2}$-wave 
superconductivity based on half-Skyrmions.
In \sref{sec_pseudogap}, we describe a pseudogap behavior
in the half-Skyrmion system.
It is shown that the energy dispersion of a composite particle
of a half-Skyrmion and doped hole
leads to an arc-like Fermi surface.

\section{Review of high-temperature superconductivity}
\label{sec_high_Tc}
\begin{figure}[ht]%
\begin{center}
  \parbox{2.1in}{\epsfig{figure=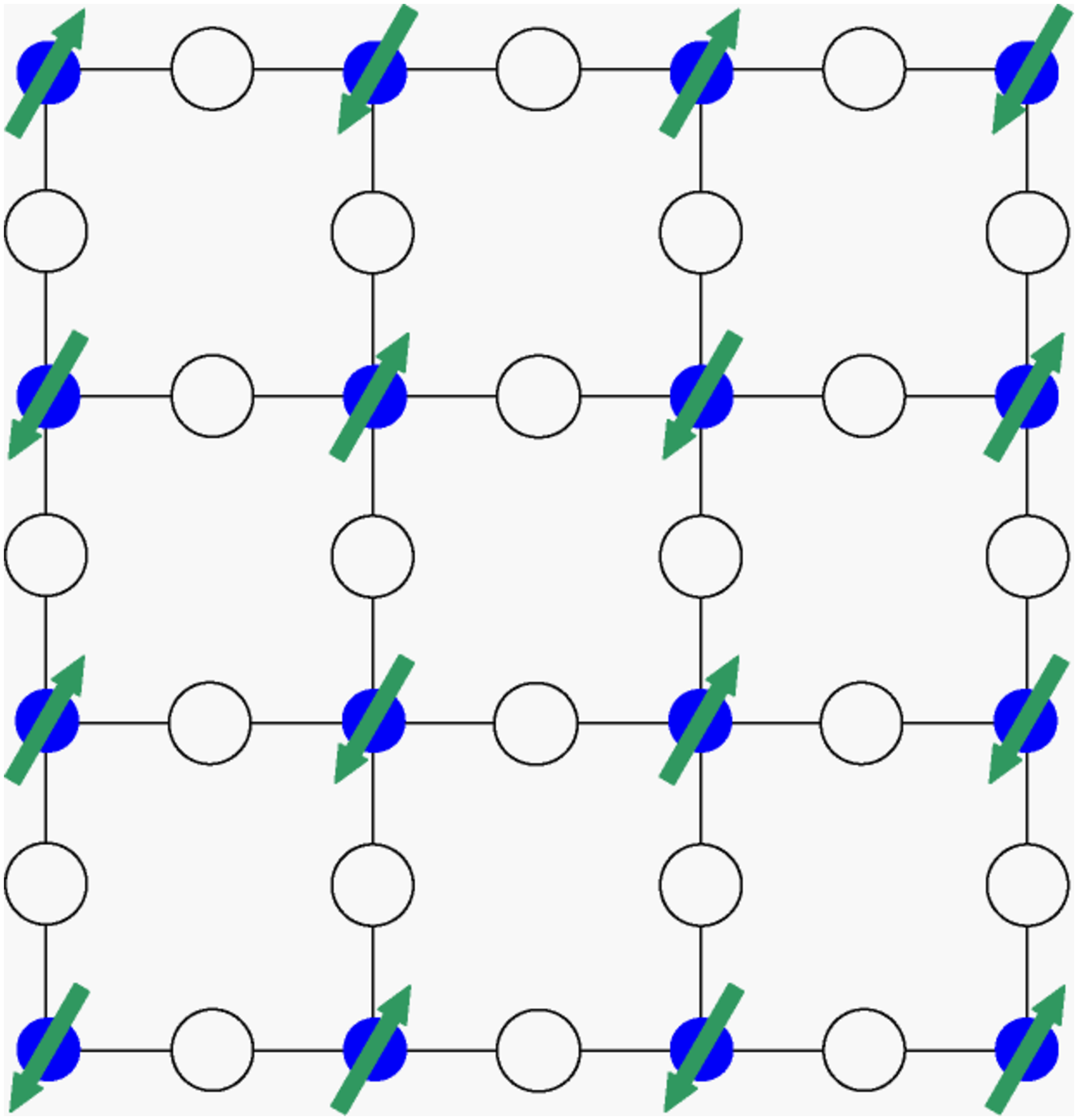,width=2in}\figsubcap{a}}
  \hspace*{4pt}
  \parbox{2.1in}{\epsfig{figure=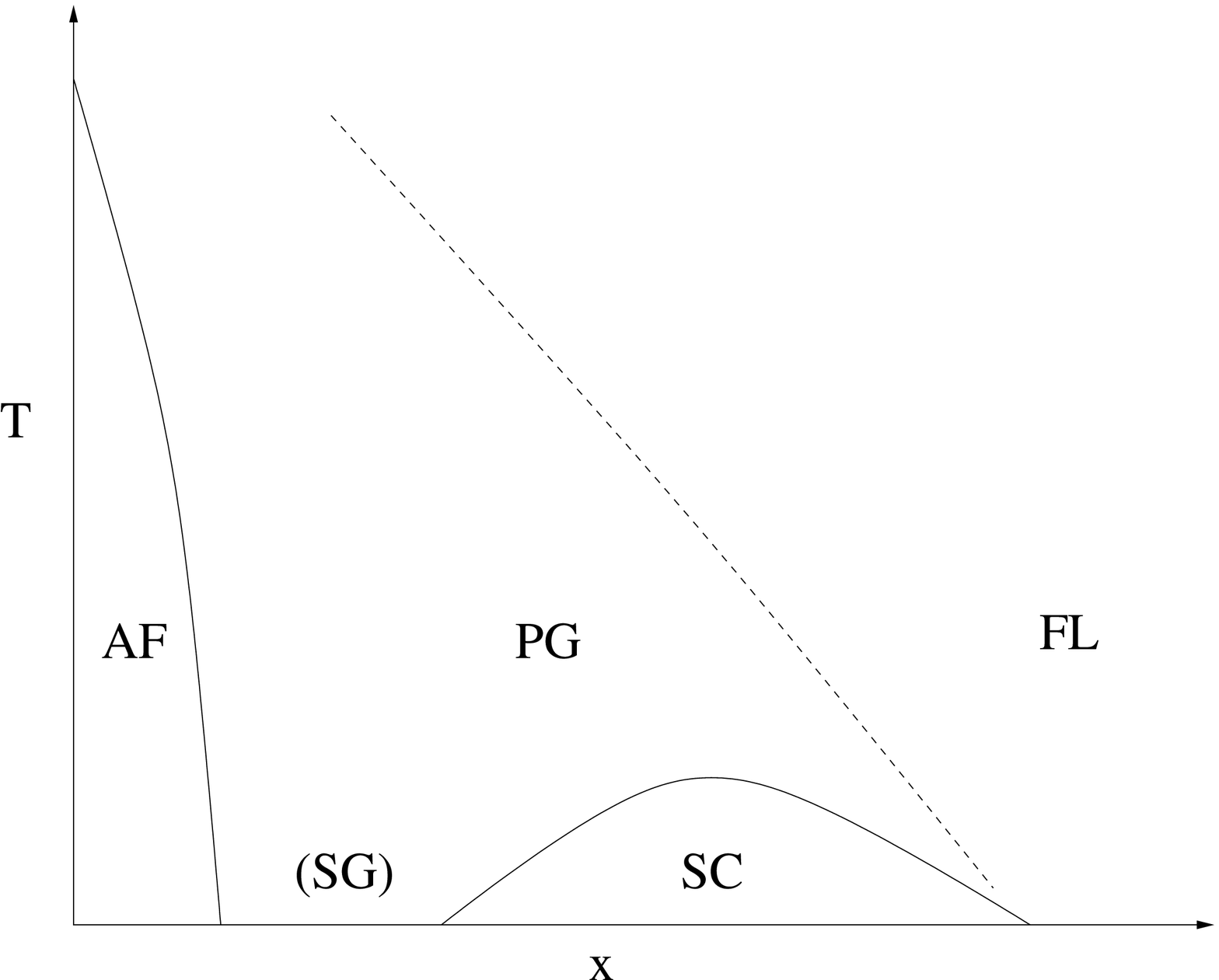,width=2in}\figsubcap{b}}
  \caption{
  (a) Two-dimensional CuO$_2$ plane. 
  Filled circles represent copper ions. 
  Open circles represent oxygen ions.
  Arrows are localized spin moments at each copper site.
  In the ground state, those spins have anti-ferromagnetic
  long-range order.
  (b) A schematic phase diagram of the high-temperature superconductors.
  The horizontal axis represents the doped hole concentration
  and the vertical axis represents temperature.
  AF indicates antiferromagnetic long-range order
	and SC indicates superconductivity.
  Below a characteristic temperature curve denoted by the dashed line, 
	the system shows a pseudogap behavior (denoted by PG).
  SG and FL indicate spin-glass-like state and Fermi liquid state,
 	respectively.
   }  
\label{fig_cuo2_pd}
\end{center}
\end{figure}

Although there are a number of high-temperature superconductors,
the essential structure is the CuO$_2$ plane.
Material differences arise from an insulating layer 
sandwiched by CuO$_2$ planes.\cite{ImadaFujimoriTokura1998}
In the parent compound, nine electrons occupy 3$d$ orbitals at
each copper ion.
In the hole picture, there is one hole at each copper site.
The hole band is half-filled but
the system is an insulator because of a strong Coulomb repulsion.
The system is well described by the spin $S=1/2$ antiferromagnetic
Heisenberg model on the square lattice with the superexchange 
interaction $J\simeq 1500 \mathrm{K}$.\cite{Manousakis1991}
Experimentally and theoretically it is established that
the ground state is an antiferromagnetic long-range ordered state.\cite{Manousakis1991}
The structure of the CuO$_2$ plane 
and the arrangement of spins at copper sites
in the undoped system are schematically shown in \fref{fig_cuo2_pd}(a).

This antiferromagnetic long-range order is rapidly suppressed 
by hole doping.
In fact only $2-3\%$ doping concentration is enough to kill 
antiferromagnetic long-range order.
This critical hole concentration is much lower than the percolation 
limit of $\sim 40\%$.
High-temperature superconductivity occurs by introducing about $0.05$ 
holes per copper ion.
A schematic phase diagram is shown in \fref{fig_cuo2_pd}(b).
(In this review we focus on the hole doped system
and do not discuss the electron doped system.)

In the high-temperature superconductors anomalous behaviors
are observed in physical quantities for temperatures above the 
superconducting transition temperature, $T_c$.\cite{Timusk1999}
The phenomenon is called pseudogap.
The Fermi surface observed by angle-resolved photoemission
spectroscopy (ARPES) in the underdoped regime
is a truncated, arc-like Fermi surface.\cite{Norman1998}
(See for a review, Ref.~\refcite{ShenRMP}.)
In scanning tunneling spectroscopy, a gap like feature appears
below the pseudogap temperature $T^*$ which is higher than $T_c$.\cite{Renner1998}
For temperatures below $T^*$ gap-like behaviors are observed
in NMR, transport coefficients, and optical conductivity.
(See for a review, Ref.~\refcite{Timusk1999}.)

In the doped system,
because of the strong Coulomb repulsion 
at each copper site\cite{ImadaFujimoriTokura1998} doped holes occupy 
oxygen $p$-orbitals.
Hole spins interact with copper site spins with 
strong antiferromagnetic Kondo interaction.
Because Kondo interaction coupling, $J_K$,
is much larger than $J$ and hole hopping matrix elements,
there is correlation of forming a singlet pair 
called the Zhang-Rice singlet.\cite{ZhangRice1988}

The strong $J_K$ limit leads to 
the $t$-$J$ model.\cite{ZhangRice1988}
The $t$-$J$ model has been studied extensively.
(See for a review, Ref.~\refcite{LeeNagaosaWen2006}.)
In the $t$-$J$ model, double occupancy is projected out.
One way to deal with this constraint is to use
slave-particle formulations.
Based on the resonating valence bond picture proposed by 
Anderson\cite{Anderson1987},
a spin-charge separation scenario has been applied 
to the physics of high-temperature superconductors.\cite{LeeNagaosaWen2006}
From various physical view points
different theories have been proposed.
There is a view in which incommensurate spin correlations
observed in neutron scattering
are associated with stripe order.
(See for a reivew, Ref.~\refcite{Kivelson2003}.)
Chakravarty {\it et al}. proposed 
d-density wave order as competing order against
superconductivity to explain the pseudogap phenomenon.\cite{Chakravarty2001}
The half-Skyrmion theory has some connection with these theories
which will be discussed later.

\section{Single hole doped system}
\label{sec_single_hole}
The high-temperature superconductors are characterized by
a rich phase diagram shown in \fref{fig_cuo2_pd}(b).
Remarkably this phase diagram is essentially controlled by 
a single parameter $x$, the doped hole concentration.
Therefore, to understand the physics of 
high-temperature superconductivity
it is necessary to establish how to describe doped holes.
Here we consider a half-Skyrmion spin texture created by 
a doped hole
in an antiferromagnetically correlated spins.

As the simplest model we consider the single hole doped system.
As stated in the previous section, the undoped system is described by
the antiferromagnetic Heisenberg model on the square lattice,
\be
H=J\sum_{\langle i,j \rangle} {\bf S}_i \cdot {\bf S}_j,
\ee
where the summation is taken over the nearest neighbor sites
and the vector ${\bf S}_i$ describes a spin $S=1/2$ at site $i$.
Theoretically and experimentally  it has been established
that the ground state is the antiferromagnetic 
long-range ordered state.\cite{Manousakis1991}
A convenient description of the state is obtained by introducing
Schwinger bosons\cite{Arovas1988} and then describing 
the long-range ordered state in terms of 
a Bose-Einstein condensate of those Schwinger bosons.
In the Schwinger boson theory the spin ${\bf S}_i$ is represented by 
\[
{\bf{S}}_i  = \frac{1}{2}\left( {\begin{array}{*{20}c}
   {\zeta_{i \uparrow }^\dag  } & {\zeta_{i \downarrow }^\dag  }  \\
\end{array}} \right) {\boldsymbol \sigma}  \left( {\begin{array}{*{20}c}
   {\zeta_{i \uparrow } }  \\
   {\zeta_{i \downarrow } }  \\
\end{array}} \right),
\]
where the components of the vector 
${\boldsymbol \sigma} = \left( \sigma_x, \sigma_y, \sigma_z \right)$ 
are Pauli spin matrices.
To describe the spin $S=1/2$, the Schwinger bosons must 
satisfy the constraint,
$\sum\limits_{\sigma  =  \uparrow , \downarrow } 
{\zeta_{j\sigma }^\dag  \zeta_{j\sigma } }  = 1$.
We introduce a mean field 
$A_{ij}=\left\langle {\zeta_{i \uparrow } \zeta_{j \downarrow }  
- \zeta_{i \downarrow } \zeta_{j \uparrow } } \right\rangle$
and introduce a Lagrange multiplier $\lambda_j$ to impose
the constraint.
In the Schwinger boson mean field theory\cite{Arovas1988},
we assume uniform values for these quantities as 
$A_{ij}=A$ and $\lambda_j=\lambda$.
The energy dispersion of Schwinger bosons is given by 
$\omega_{\bf k} = \sqrt{\lambda^2-4J^2 A^2 \gamma_{\bf k}^2}$
with $\gamma_{\bf k}=(\sin k_x + \sin k_y)/2$.
In the ground state, $\lambda=2JA$.
Bose-Einstein condensation 
occurs \cite{Hirsch1989,Yoshioka1989,Raykin1993}  
at ${\bf k}=(\pm \pi/2, \pm \pi/2)$.

Now we consider a hole introduced in the system.
The strong interaction between the doped hole spin and copper 
site spins leads to correlation of forming
a Zhang-Rice spin singlet\cite{ZhangRice1988}
as mentioned in the previous section.
If the singlet is formed, then the Bose-Einstein condensate
of the Schwinger bosons is suppressed around the doped hole position.
Generally if the condensate is suppressed at some point
in two-dimensional space, 
then a vortex is formed around that point.
The vortex solution is found by solving
the Gross-Pitaevskii equation.\cite{FetterWalecka}
For the Schwinger bosons, the vortex turns out to be
a half-Skyrmion as shown below.

For the description of the half-Skyrmion, it is convenient 
to use the non-linear sigma model.\cite{Chakravarty1988}
Low-energy physics of the antiferromagnetic Heisenberg model
is well described by the non-linear sigma model\cite{Chakravarty1988},
\be
S=\frac{\rho_{\mathrm{s}}}{2} \int_0^{\left(k_B T \right)^{-1}} d\tau 
\int d^2 {\bf r} 
\left[
\frac{1}{c_{\mathrm{sw}}^2} 
\left( \frac{\partial \bf n}{\partial \tau} \right)^2
+
\left( \nabla {\bf n} \right)^2
\right],
\ee
where $\rho_{\mathrm{s}}$ is the spin stiffness and $c_{\mathrm{sw}}$ is the antiferromagnetic
spin-wave velocity.
(Hereafter we use units in which $\hbar=1$.)
The unit vector ${\bf n}$ represents the staggered moment
and $\tau$ is the imaginary time.

In order to describe the correlation of forming a Zhang-Rice singlet pair
between doped hole spins and copper site spins,
one has to be careful about its description.
Obviously forming a static singlet state which is realized in the
$J_K \rightarrow \infty$ limit does not work.
Because such a simple singlet state contradicts with the rapid 
suppression of antiferromagnetic long-range order by hole doping.
If static singlet states are formed, then sites occupied by
singlets do not interact with the other spins at all.
The situation is similar to site dilution,
and suppression of magnetic long-range order is described 
by the percolation theory.
In other words, considering a strongly localized wave function
of a doped hole at a copper site is not realistic.
We need to consider a hole wave function extending over some area
so that the doped hole spin interacts with the other spins.
In fact, numerical diagonalization studies of the $t$-$J$ model
show a Skyrmion-like spin texture \cite{Gooding1991}
when a hole motion is restricted to one plaquette.
A similar situation may be realized in Li-doped system
as discussed in Ref.~\refcite{Haas1996}.

To include the effect of the interaction with the other spins,
we formulate the correlation of forming a Zhang-Rice singlet
in the following way.
The spin singlet wave function of a copper site spin and 
a hole spin is described by
\[
\frac{1}{{\sqrt 2 }}
\left( 
   \left|  \uparrow  \right \rangle_{\mathrm{h}} 
   \left|  \downarrow  \right \rangle_{\mathrm{Cu}}  
 - \left|  \downarrow  \right \rangle_{\mathrm{h}}  
   \left|  \uparrow  \right \rangle_{\mathrm{Cu}}
\right).
\]
This wave function has the form of superposition of 
the hole spin-up and copper spin-down state,
$\left|  \uparrow  \right\rangle _{\mathrm{h}} 
\left|  \downarrow  \right\rangle _{\mathrm{Cu}}$
and the hole spin-down and copper spin-up state,
$\left|  \downarrow  \right\rangle _{\mathrm{h}} \left|  \uparrow  \right\rangle _{\mathrm{Cu}}$.
In order to include the interaction effect,
we consider these states separately and 
construct superposition of them.
We assume that the spin state at site $j$ is spin-up
before the introduction of a doped hole.
Under this assumption, the spin-up state does not change 
directions of the neighboring spins. 
So the system is uniform for the staggered spin ${\bf n}$.
By contrast, the spin-down state at site $j$
creates a Skyrmion spin texture characterized by a topological charge,
\[
Q = \frac{1}{{8\pi }}\int {d^2 {\bf{r}}} 
\varepsilon_{\alpha \beta}
\,{\bf{n}}\left( {\bf{r}} \right) \cdot 
\left[ {\partial _{\alpha} {\bf{n}}\left( {\bf{r}} \right) 
\times \partial _{\beta} {\bf{n}}\left( {\bf{r}} \right)} \right],
\]
where $\varepsilon_{xx}=\varepsilon_{yy}=0$
and $\varepsilon_{xy}=-\varepsilon_{yx}=1$.
Following Ref.~\refcite{Belavin1975}, the Skyrmion solution
is found by making use of an inequality
\[
\int {d^2 {\bf{r}}} \left[ {\partial _\alpha  {\bf{n}} 
\pm \varepsilon _{\alpha \beta } 
\left( {{\bf{n}} \times \partial _\beta  {\bf{n}}} \right)} \right]^2  \ge 0.
\]
The classical energy satisfies,
\[
E = \frac{{\rho _{\mathrm{s}} }}{2}\int {d^2 {\bf{r}}} 
\left( {\nabla {\bf{n}}} \right)^2  \ge 4\pi \rho _{\mathrm{s}} Q.
\]
The equality holds if and only if 
\be
\partial _\alpha  {\bf{n}} \pm \varepsilon _{\alpha \beta } 
\left( {{\bf{n}} \times \partial _\beta  {\bf{n}}} \right) = 0.
\label{eq_equality_cond}
\ee
This equation is rewritten in a simple form. 
If we introduce
\[
w = \frac{{n_x  + in_y }}{{1 - n_z }},
\]
then \eref{eq_equality_cond} is rewritten as
\[
\left( {\partial _x  \mp i\partial _y } \right)w = 0.
\]
This equation is the Caucy-Riemann equation.
Noting
\[
{\bf{n}} = \left( {\frac{{2{\mathop{\rm Re}\nolimits} w}}
{{\left| w \right|^2  + 1}},\frac{{2{\mathop{\rm Im}\nolimits} w}}
{{\left| w \right|^2  + 1}},\frac{{\left| w \right|^2  - 1}}
{{\left| w \right|^2  + 1}}} \right),
\]
the solutions satisfying the boundary condition 
${\bf n} ({\bf r}_j) = -{\hat z}$
and ${\bf n} ({\bf r}\rightarrow \infty) = {\hat z}$ 
with ${\hat z}$ the unit vector along the $z$-axis
are the Skyrmion spin texture,
\[
{\bf{n}}\left( {\bf{r}} \right) = \left( 
{\frac{{2\eta x}}{{r^2  + \eta ^2 }},\frac{{2\eta y}}{{r^2  + \eta ^2 }},
\frac{{r^2  - \eta ^2 }}{{r^2  + \eta ^2 }}} \right),
\]
with $Q=1$ and the anti-Skyrmion spin texture,
\[
{\bf{n}}\left( {\bf{r}} \right) = \left( 
{\frac{{2\eta x}}{{r^2  + \eta ^2 }}, -\frac{{2\eta y}}{{r^2  + \eta ^2 }},
\frac{{r^2  - \eta ^2 }}{{r^2  + \eta ^2 }}} \right),
\]
with $Q=-1$.

\begin{figure}[ht]%
\begin{center}
  \parbox{2.1in}{\epsfig{figure=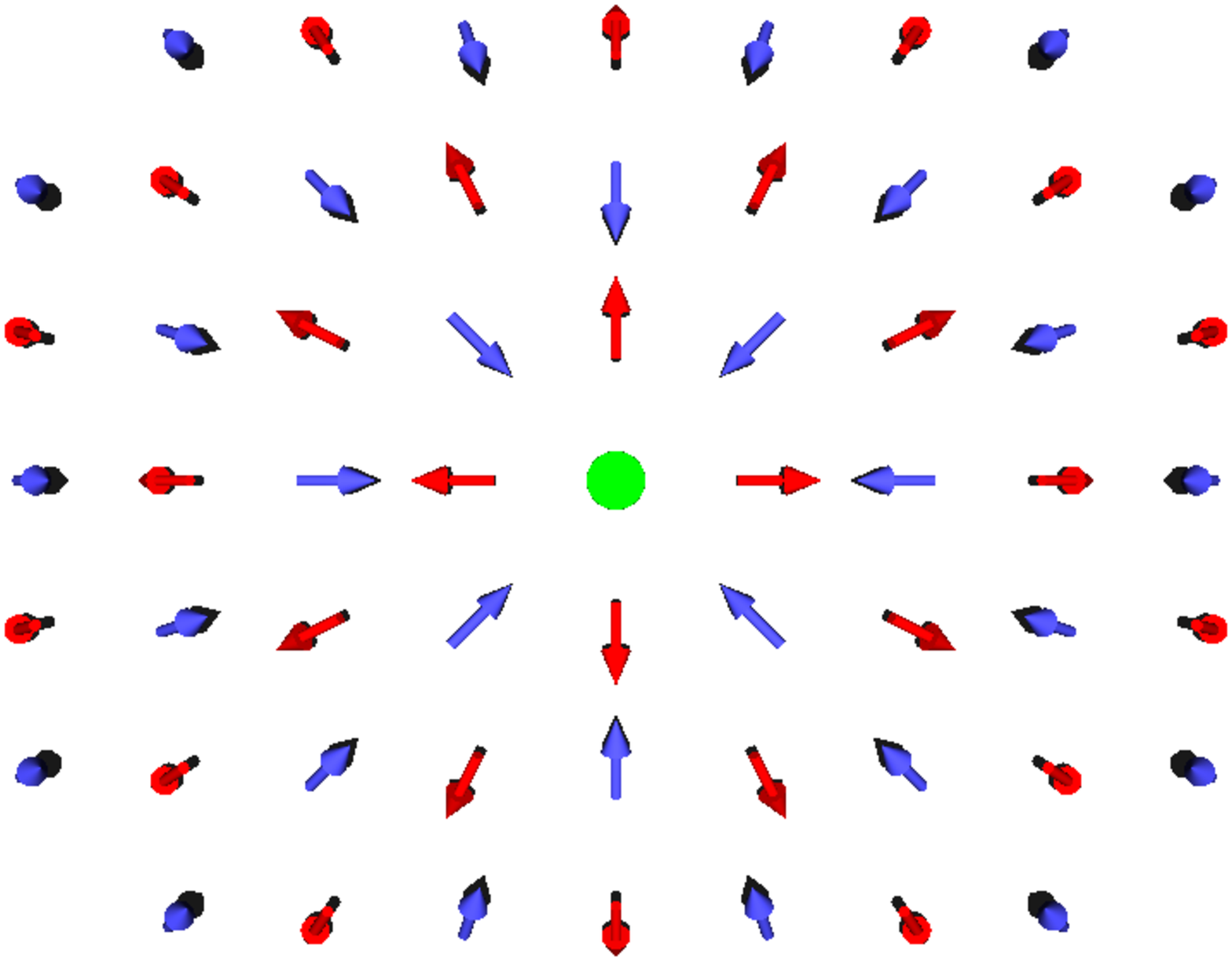,width=2in}\figsubcap{a}}
  \hspace*{4pt}
  \parbox{2.1in}{\epsfig{figure=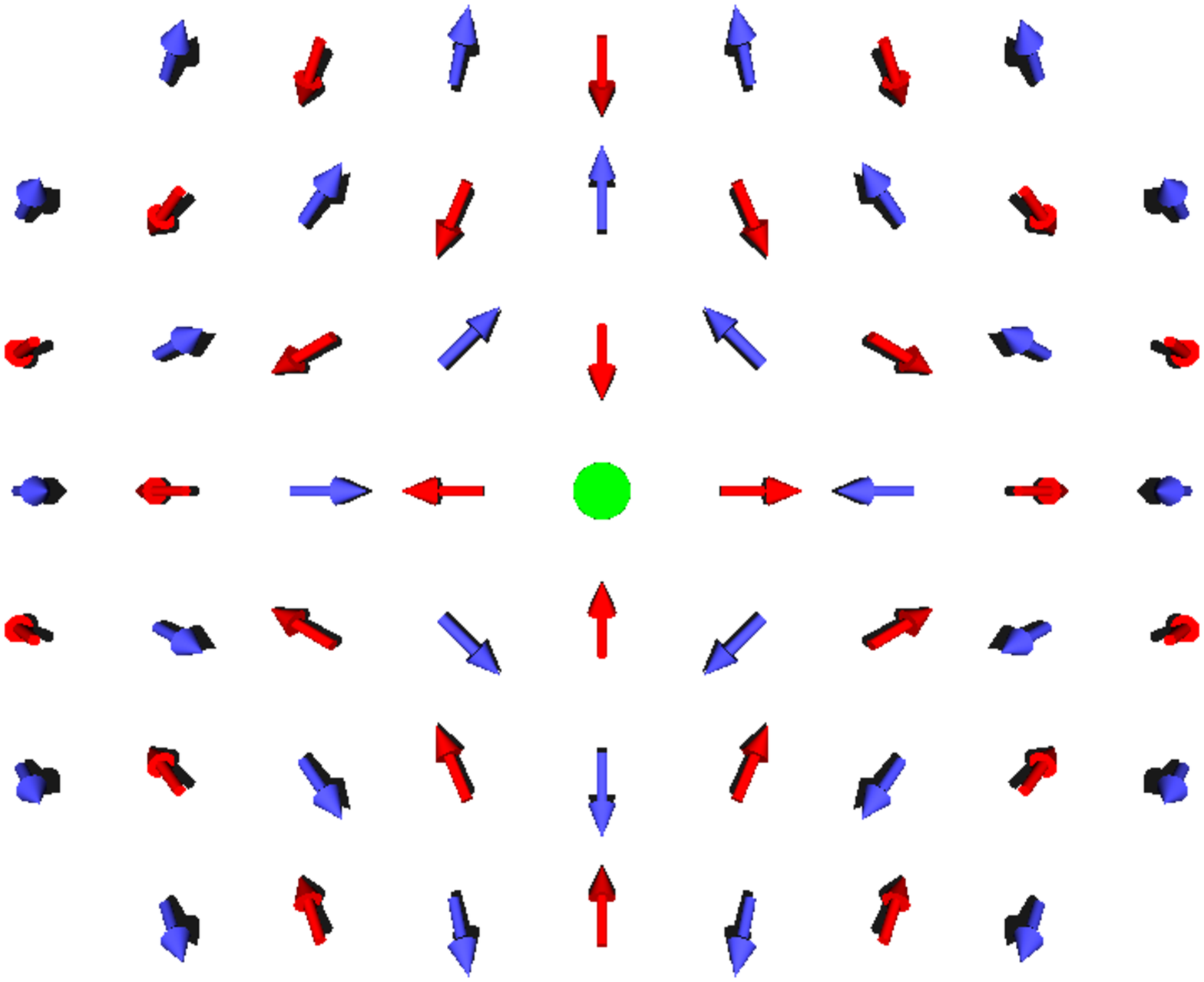,width=2in}\figsubcap{b}}
  \caption{(a) Half-Skyrmion spin texture.
  Arrows represent the directions of the spin at the copper sites.
  Neighboring spins are almost anti-parallel because of the antiferromagnetic
  correlations.
  Filled circle at the center denotes the core of the half-Skyrmion.
  (b) Anti-half-Skyrmion spin texture.}  
\label{fig_hs}
\end{center}
\end{figure}

Now we consider superposition of the uniform state and
the Skyrmion state. 
Although superposition of the two spin 
configurations is not the solution of the field
equation, these solutions suggest that the resulting
spin configuration is characterized by a topological charge $Q$
with $0<|Q|<1$. 
The value of $Q$ is determined by making use of the fact that 
the antiferromagnetic long-range ordered state is described by
Bose-Einstein condensation of Schwinger bosons.
In the CP$^1$ representation of the non-linear sigma model\cite{Rajaraman},
the U(1) gauge field is introduced by
\be
\alpha _\mu   =  - i
\sum_{\sigma=\uparrow, \downarrow}
z _\sigma ^* \partial _\mu  z _\sigma
\label{eq_gauge_field}
\ee
where the complex field $z_{\sigma}$ is defined through
\be
{\bf{n}} = \left( {\begin{array}{*{20}c}
   {z _ \uparrow ^* } & {z _ \downarrow ^* }  \\
\end{array}} \right) \sigma  \left( {\begin{array}{*{20}c}
   {z _ \uparrow  }  \\
   {z _ \downarrow  }  \\
\end{array}} \right).
\ee
In terms of the gauge field $\alpha_{\mu}$, 
the topological charge $Q$ is rewritten as
\[
Q = \int {\frac{{d^2 {\bf{r}}}}{{2\pi }}
\left( {\partial _x \alpha _y  - \partial _y \alpha _x } \right)}.
\]
From this expression, we see that a spin configuration with
$Q$ corresponds to the flux $2\pi Q$ in the condensate of the
Schwinger bosons. 
On the other hand, each Schwinger boson carries the spin $S=1/2$.
However, there are no $S=1/2$ excitations 
in the low-energy excitation spectrum.
Low-lying excitations are antiferromagnetic spin waves
which carry the spin one.
Therefore, all Schwinger bosons are paired
and the flux quantum is $\pi$ similarly to conventional
BCS superconductors.\cite{Ng1999}
The flux value is not arbitrary and $Q$ must be in the form 
of $Q=n/2$, with $n$ an integer. 
Taking into account the constraint $0<|Q|<1$,
we may conclude $|Q|=1/2$.\cite{Morinari2005}
The spin texture with $|Q|=1/2$ is called half-Skrymion spin texture
because the topological charge is half of the Skyrmion spin texture.
The half-Skyrmion spin texture and the anti-half-Skyrmion spin texture 
are schematically shown in \fref{fig_hs}.

Moving half-Skyrmion spin texture is obtained by
applying Lorentz boost on the static solution above
by making use of the Lorentz invariance 
of the non-linear sigma model.\cite{Rajaraman}
The energy dispersion is
\be
E_{\bf k} = \sqrt{c_{\mathrm{sw}}^2 k^2 + E_0^2},
\ee
where $E_0=2\pi \rho_{\mathrm{s}}$ is the half-Skyrmion creation energy.
On the square lattice the dispersion is transformed into
\be
E_{\bf k} = \sqrt{c_{\mathrm{sw}}^2 
\left( 
\cos^2 k_x + \cos^2 k_y
\right) + E_0^2}.
\label{eq_hs_disp}
\ee
Note that the lowest energy states are at $(\pm \pi/2, \pm \pi/2)$
because the Schwinger bosons are gapless at those points in the 
antiferromagnetic long-range ordered state.
The half-Skyrmion spin texture are mainly formed 
by Schwinger bosons around those points.

Now we compare the half-Skyrmion excitation spectrum
with the ARPES result in the undoped system.
The excitation spectrum \eref{eq_hs_disp} is 
qualitatively in good agreement with excitation spectrum
obtained by Wells {\it et al}.\cite{Wells1995}
The parameters $c_{\mathrm{sw}}$ and $E_0$ are determined from
the values for the Heisenberg antiferromagnet. 
We use the renormalized factors $Zc=1.17$ and $Z_{\rho}=0.72$, 
which are estimated from quantum Monte Carlo simulations
\cite{Beard1998,Kim1998}
and a series expansion technique.\cite{Singh1989}
Using these values, we find that the bandwidth is $1.5J$ 
and $E_0=1.1J$. 
The experimentally estimated bandwidth by Wells et al. is
$2.2J$.
From the fitting of the experimental data assuming \eref{eq_hs_disp},
we find $E_0 \sim J$.
\begin{figure}
\centerline{\psfig{file=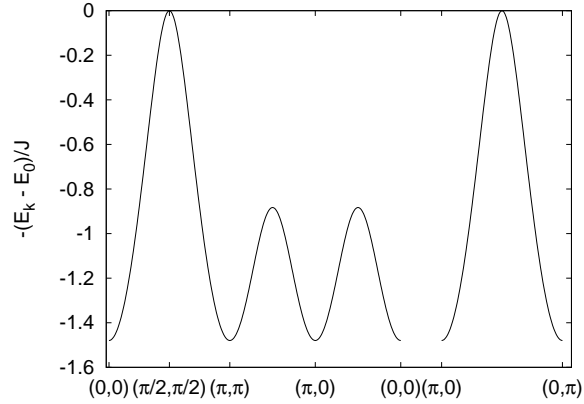,width=8.0cm}}
\caption{
The half-Skyrmion dispersion.
Horizontal axis represents positions in the Brillouin zone.
}
\label{fig_hs_disp}
\end{figure}

In the undoped system, anomalously broad line shapes are observed by
ARPES.
Line shape broadening is associated with scattering of
excitations by fluctuation modes.
In the half-Skyrmion theory, half-Skymions couple to
spin-wave excitations.
Describing those spin wave excitations in terms of 
the gauge field fluctuations
line shape broadening is studied by applying
a strong coupling analysis.\cite{Morinari2008scc}
The width of the broadening is 
in good agreement with the experiment.

\section{Multi half-Skyrmion configurations}
\label{sec_multi_hs_config}
In the previous section, the single half-Skyrmion
has been considered.
The most important physical quantity carried by 
the half-Skyrmion is the topological charge.
The topological charge density, which is defined in the continuum as
\[
q_c \left( {\bf{r}} \right) = \frac{1}{{4\pi }}{\bf{n}}
\left( {\bf{r}} \right) \cdot \left[ {\partial _x {\bf{n}}
\left( {\bf{r}} \right) \times \partial _y {\bf{n}}
\left( {\bf{r}} \right)} \right],
\]
has the following form on the lattice,
\[
q_c \left( {x_j,y_j} \right) = \frac{1}{{16\pi }}{\bf{n}}
\left( {x_j,y_j} \right) \cdot \left[ {{\bf{n}}
\left( {x_j + 1,y_j} \right) - {\bf{n}}\left( {x_j - 1,y_j} \right)} \right] 
\times \left[ {{\bf{n}}\left( {x_j,y_j + 1} \right) - 
{\bf{n}}\left( {x_j,y_j - 1} \right)} \right].
\]
In the single half-Skyrmion state,
$q_c(x_j,y_j)$ has a peak around the half-Skyrmion position,
and vanishes at infinity.
The integration of $q_c({\bf r})$ 
leads to the quantized value $Q=\pm1/2$.

\begin{figure}
\centerline{\psfig{file=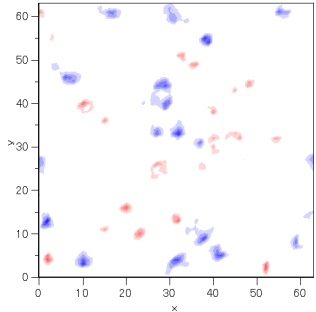,width=4.2cm}}
\caption{
Topological charge density distribution in real space at 
the doping concentration $x=0.107$ on a $64 \times 64$ lattice.
Positive values are shown in red and negative values are shown
in blue.
}
\label{fig_chirality}
\end{figure}

If there are many half-Skyrmions, do half-Skyrmions
keep topological charge?
In order to answer this question, 
we carry out a simple numerical simulation.
First, we put either an XY-spin-vortex or an anti-XY-spin-vortex randomly.
(A similar numerical simulation is discussed in Ref.~\refcite{Berciu2004}.)
A multi-XY-spin-vortex configuration is defined by
\bea
n_x ({\bf r}) &=& \sum\limits_j {q_j \frac{{x - x_j }}{{\left( {x - x_j } \right)^2  
+ \left( {y - y_j } \right)^2 }}}, \\ 
n_y ({\bf r}) &=& \sum\limits_j {q_j \frac{{y - y_j }}{{\left( {x - x_j } \right)^2  
+ \left( {y - y_j } \right)^2 }}},
\eea
where $q_j=+1$ for XY-spin-vortices and $q_j=-1$ 
for anti-XY-spin-vortices.
A doped hole is sitting at each spin-vortex position $(x_j,y_j)$,
and ${\bf n}(x_j,y_j)=0$.
Then, a random number which ranges from $-0.1$ to $0.1$
is assigned to the $z$-component of 
the vector ${\bf n}(x_i,y_i)$
except for the nearest neighbor sites 
$(x_i\pm 1, y_i)$ and $(x_i,y_i\pm 1)$.
After that, the equilibrium configuration of 
the vectors ${\bf n}(x_i,y_i)$ is obtained by the relaxation method.
At site $(x_{\ell},y_{\ell})$, ${\bf n}(x_{\ell},y_{\ell})$ is updated by 
\[
{\bf{n}}\left( {x_{\ell},y_{\ell}} \right) = 
\frac{1}{4}\left[ {{\bf{n}}\left( {x_{\ell} + 1, y_{\ell}} \right) 
+ {\bf{n}}\left( {x_{\ell} - 1,y_{\ell}} \right) 
+ {\bf{n}}\left( {x_{\ell}, y_{\ell} + 1} \right) 
+ {\bf{n}}\left( {x_{\ell},y_{\ell} - 1} \right)} \right].
\]
The constraint $|{\bf n}(x_{\ell},y_{\ell})|=1$ is imposed by taking 
the normalization after the update.
This update procedure is carried out over all lattice sites
except for the hole positions $(x_j,y_j)$ and 
its nearest neighbor sites,
$(x_j \pm 1, y_j)$ and $(x_j, y_j \pm 1)$.
The resulting converged state is an approximate state for
a multi-half-Skyrmion-anti-half-Skyrmion configuration.
As an example, \fref{fig_chirality}
shows the topological charge density distribution 
at the doping concentration $x=0.107$.
There are regions where topological charge density
is non-zero.
Positive (negative) topological charge density region
is associated with half-Skyrmions (anti-half-Skyrmions).
Non-vanishing distribution patterns are 
observed for $x<x_c$ with $x_c \sim 0.30$.
For $x>x_c$, half-Skyrmions and anti-half-Skyrmions are heavily overlapped.
As a result topological charges are cancelled out.
Therefore, above $x_c$ the topological nature of half-Skyrmions 
is lost.
(For related discussions about the effect of thermally excited 
skyrmions and hole induced skyrmions, 
see Refs.~\refcite{Belov1998,Timm2000}.)

Now we discuss magnetic properties of multi-half-Skyrmion
configurations.
The magnetic correlation is investigated by the static 
magnetic structure factor,
\[
S\left( {\bf{q}} \right) = \sum\limits_{\alpha ,\beta } 
{\left( {\delta _{\alpha \beta }  
- \frac{{q_\alpha  q_\beta  }}{{q^2 }}} \right)
S_{\alpha \beta } \left( {\bf{q}} \right)}.
\]
Here
\[
S_{\alpha \beta } \left( {\bf{q}} \right) 
= \frac{1}{N}\sum\limits_{i,j} {e^{i{\bf{q}} 
\cdot \left( {{\bf{R}}_i  - {\bf{R}}_j } \right)} S_{i\alpha } S_{j\beta } }.
\]
$S({\bf q})$ is measured by neutron scattering experiments.
Introducing,
\[
{\bf{S}}\left( {\bf{q}} \right) = 
\sum\limits_j {e^{i{\bf{q}} \cdot {\bf{R}}_j } {\bf{S}}_j },
\]
$S({\bf q})$ is rewritten as
\[
S\left( {\bf{q}} \right) = 
\frac{1}{N}\left[ {\left| {{\bf{S}}
\left( {\bf{q}} \right)} \right|^2  
- \frac{1}{{q^2 }}\left| {{\bf{q}} \cdot 
{\bf{S}}\left( {\bf{q}} \right)} \right|^2 } \right].
\]
If there is antiferromagnetic long-range order, then
$S({\bf q})$ has a peak at ${\bf q}=(\pi,\pi) \equiv {\bf Q}$,
and the peak height is proportional to the number of 
lattice sites.
From numerical simulations above, we find that
$S({\bf q})$ shows incommensurate peaks 
at positions shifted from ${\bf q}={\bf Q}$.
Furthermore, we find that around $x=0.10$
the maximum peak height is on the order of the square root 
of the number of lattice sites.
Therefore, the magnetic long-range order disappears
around that doping concentration.

\begin{figure}[ht]%
\begin{center}
  \parbox{2.1in}{\epsfig{figure=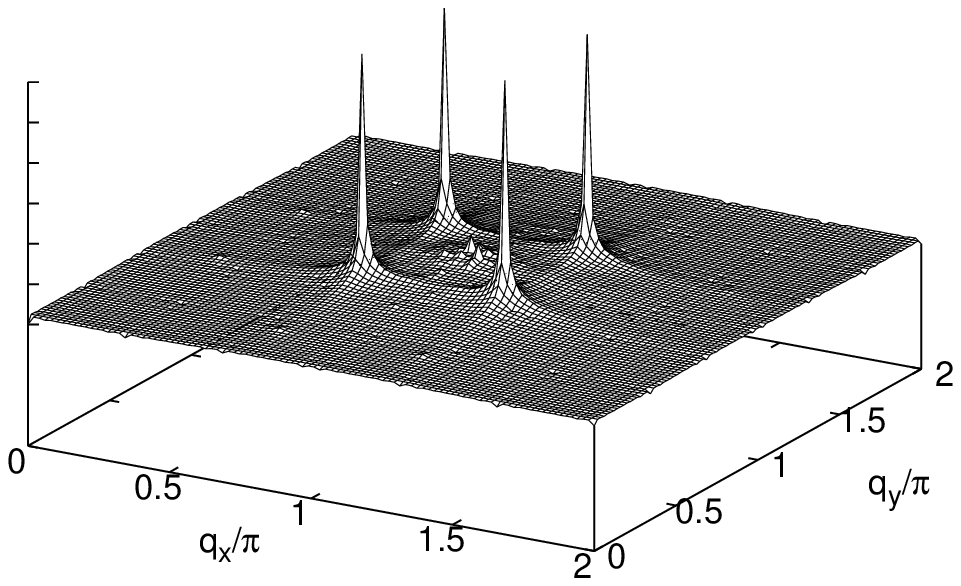,width=2in}\figsubcap{a}}
  \hspace*{4pt}
  \parbox{2.1in}{\epsfig{figure=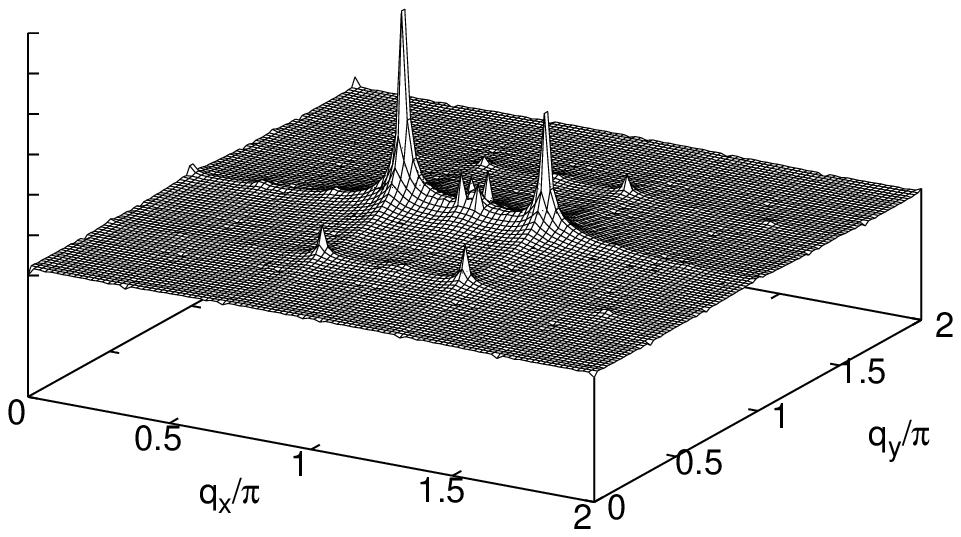,width=2in}\figsubcap{b}}
  \caption{
  (a) Magnetic structure factor in a roughly antiferromagnetically ordered
	half-Skyrmions and anti-half-Skyrmions.
  (b) Magnetic structure factor in a stripe-like configuration
	of half-Skyrmions and anti-half-Skyrmions.
  }
  \label{fig_Sq}
\end{center}
\end{figure}

The physical origin of the incommensurate peaks
is found by studying a regular configuration of half-Skyrmions.
Taking a vortex-anti-vortex configuration given by 
$q_{(x_j,y_j)}=(-1)^{x_j+y_j}$,
an approximate "antiferromagnetic" configuration of
half-Skyrmions and anti-half-Skyrmions is obtained by the numerical simulation.
\Fref{fig_Sq}(a) shows the magnetic structure factor of the resulting state
at $x=0.0625$ on a $64 \times 64$ lattice.
The incommensulate peaks are found at
$(\pi (1 \pm 2 \delta), \pi (1 \pm 2 \delta))$
with $\delta=0.125$.
These peaks are associated with the superlattice formed by
half-Skyrmions and anti-half-Skyrmions.
A stripe order case is shown in \ref{fig_Sq}(b)
which is obtained by taking 
a vortex-anti-vortex configuration with $q_{(x_j,y_j)}=(-1)^{x_j}$.
The dominant incommensulate peaks are found at
$(\pi (1 \pm 2 \delta), \pi)$ with $\delta=0.125$.

Experimentally neutron scattering experiments show 
incommensurate peaks at
${\bf q}=(\pi (1 \pm 2\delta), \pi (1 \pm 2\delta))$
for $x<0.05$ and 
${\bf q}=(\pi (1 \pm 2\delta), \pi )$
and 
${\bf q}=(\pi, \pi (1 \pm 2 \delta))$ for $x>0.05$.
\cite{Cheong1991,Yamada1998,Matsuda2000}
As shown above, such incommensurate peaks are 
found in some configurations of half-Skyrmions and anti-half-Skyrmions.
However, there is a quantitative difference.
Experimentally it is found that $\delta \simeq x$.\cite{Yamada1998,Matsuda2000}
In order to explain this experimental result,
it is necessary to consider
stripe-like configurations of half-Skyrmions and anti-half-Skyrmions.
To determine the stable configuration of half-Skyrmions and anti-half-Skyrmions,
we need to take into account the interaction between half-Skyrmions,
which is not included in the numerical simulation above.
Determination of the stable half-Skyrmion configuration
is left for future work.

\section{Mechanism of d-wave superconductivity}
\label{sec_mechanism}
In the half-Skyrmion theory, doped holes create either a half-Skyrmion 
or an anti-halr-Skyrmion at their positions.
A half-Skyrmion or an anti-half-Skyrmion is bound to 
each hole and moves together.
There are two ways to formulate the effect of 
half-Skyrmions on the doped holes.
One way is to take a doped hole and the half-Skyrmion created by the hole
as a composite particle.
This approach is formulated in the next section,
and we shall see that the theory leads to pseudogap behavior.

The other way is to include the effect of half-Skyrmions
as fields mediating interaction between doped holes.
By integrating out the half-Skyrmion degrees of freedom,
we obtain the interaction between doped holes mediated by
the half-Skyrmions.
In this section, we take this approach and show that 
the interaction leads to a $d_{x^2-y^2}$-wave Cooper pairing
between the doped holes.\cite{Morinari2006,Morinari2002}
An intuitive interpretation is also given about 
the origin of the pairing interaction based on 
a Berry phase.

The fact that each doped hole carries a half-Skyrmion
is represented by
\be
\nabla  \times \alpha  = \pi \sum\limits_{s =  \pm } 
{s\psi _s^\dag  \left( {\bf{r}} \right)\psi _s \left( {\bf{r}} \right)},
\label{eq_density_flux}
\ee
where $\alpha$ is the U(1) gauge field in the CP$^1$ model 
defined by \eref{eq_gauge_field}.
The index $s$ labels the sign of the topological charge.
$s=+$ refers to a half-Skyrmion and $s=-$ refers to an anti-half-Skyrmion.

The interaction between the doped hole current and the gauge field 
is found as follows.
Doped holes interact with the spins via a strong Kondo coupling,
\[
H_K  = J_K \sum\limits_j {{\bf{S}}_j  \cdot \left( {c_j^\dag  \sigma c_j } \right)}.
\]
Meanwhile, the doped hole motion is described by
\[
H_{\mathrm{t}}  =  - t\sum\limits_{\left\langle {i,j} \right\rangle } 
{\left( {c_i^\dag  c_j  + h.c.} \right)}.
\]
From a perturbative calculation for the tight-binding model 
describing the CuO$_2$ plane,
we find the Kondo coupling is $J_K \simeq 1 {\mathrm eV}$.
Since $J_{\mathrm K}$ is larger than $t \simeq 0.4 {\mathrm eV}$,
we first diagonalize the Kondo coupling term.
The diagonalization is carried out by the following unitary transformation,
\[
c_j  = U_j f_j,
\]
where
\[
U_j  = \left( {\begin{array}{*{20}c}
   {z_{j \uparrow } } & { - z_{j \downarrow }^* }  \\
   {z_{j \downarrow } } & {z_{j \uparrow }^* }  \\
\end{array}} \right).
\]
Under this transformation, the hopping term is
\[
H_{\mathrm{t}}  =  - t\sum\limits_{\left\langle {i,j} \right\rangle } 
{\left( {f_i^\dag  U_i^\dag  U_j f_j  + h.c.} \right)}. 
\]
Extracting the terms including the gauge field $\alpha_{\mu}$, we find
\[
H_{{\mathop{\rm int}} }  = it\sum\limits_j {\sum\limits_{\delta  = x,y} 
{f_{j + \delta }^\dag  \alpha _\delta  \sigma _z f_j } }  + h.c.
\]
(The effect of other terms is discussed in
Ref.\refcite{Shraiman1988}.)
After Fourier transforming and taking the continuum limit,
we obtain
\be
H_{{\mathop{\rm int}} }  \simeq \frac{1}{{m\Omega ^{1/2} }}
\sum\limits_{{\bf k},{\bf q}} { {\sum\limits_{\delta  = x,y} 
{\alpha _\delta  \left( {\bf q} \right) \cdot 
\left( {{\bf k} + \frac{{\bf q}}{2}} \right)_\delta  
f_{{\bf k} + {\bf q}}^\dag  \sigma _z f_{{\bf k}} } } }.
\label{eq_minc}
\ee
Here the effective mass $m$ is introduced by $t \simeq 1/2m$.

Now we derive the pairing interaction between the doped holes
from Eqs.~(\ref{eq_density_flux}) and (\ref{eq_minc})
by eliminating the gauge field.
In order to fix the gauge, we take the Coulomb gauge.
In wavevector space, we set
$\alpha _x \left( {\bf q} \right) =  - \frac{{iq_y }}{{q^2 }}\alpha 
\left( q \right)$ and
$\alpha _y \left( {\bf q} \right) = \frac{{iq_x }}{{q^2 }}\alpha \left( q \right)$.
From \eref{eq_density_flux}, we obtain
\[
\alpha \left( q \right) =  - \frac{\pi }{{\Omega ^{1/2} }}
\sum\limits_{{\bf k},s} {sf_{{\bf k}s}^\dag  f_{{\bf k} + {\bf q},s} }.
\]
Substituting this expression into \eref{eq_minc}, we obtain
\[
H_{{\mathop{\rm int}} }  \simeq  - \frac{{i\pi }}{{m\Omega }}
\sum\limits_{{\bf k},{\bf k}',{\bf q}} {\sum\limits_{s,s',\sigma ,\sigma '} 
{\frac{{k_x q_y  - k_y q_x }}{{q^2 }}s'\sigma 
f_{{\bf k}'s'\sigma '}^\dag  
f_{{\bf k}' + {\bf q},s'\sigma '} f_{{\bf k} + {\bf q},s,\sigma }^\dag  
f_{{\bf k},s,\sigma } } }.
\]
Since we are interested in a Cooper pairing, we focus on terms with
${\bf k}+{\bf k}'+{\bf q}=0$.
After symmetrizing the terms with respect to spin and half-Skyrmion indices,
we obtain
\be
H_{{\mathop{\rm int}} }  \simeq  - \frac{{i\pi }}{{m\Omega }}
\sum\limits_{{\bf{k}} \ne {\bf{k'}}} {\sum\limits_{s,s',\sigma ,\sigma '} 
{\frac{{k_x k'_y  - k_y k'_x }}{{\left( {{\bf{k}} - {\bf{k}}'} \right)^2 }}
\left( {s'\sigma  + s\sigma '} \right)
f_{{\bf k}'s'\sigma '}^\dag  f_{ - {\bf k}',s,\sigma }^\dag  
f_{ - {\bf k},s,\sigma } } } f_{{\bf k},s'\sigma '}.
\label{eq_interaction}
\ee
This interaction term leads to a pairing of holes as shown below.

Following a standard procedure,\cite{Sigrist1991}
we apply the BCS mean field theory to the interaction (\ref{eq_interaction}).
The mean field Hamiltonian reads,
\bea
H &=& \sum\limits_{{\bf{k}},s,\sigma } {\left( {\begin{array}{*{20}c}
   {f_{{\bf{k}},s,\sigma }^\dag  } & {f_{{\bf{k}}, - s, - \sigma }^\dag  } & {f_{ - {\bf{k}},s,\sigma } } & {f_{ - {\bf{k}}, - s, - \sigma } }  \\
\end{array}} \right)} 
\nonumber \\
& & \times
\left( {\begin{array}{*{20}c}
   {\xi _{\bf{k}} } & 0 & {\Delta _{{\bf{k}},s,\sigma }^{\left(  +  \right)} } & {\Delta _{{\bf{k}},s,\sigma }^{\left(  -  \right)} }  \\
   0 & {\xi _{\bf{k}} } & {\Delta _{{\bf{k}}, - s, - \sigma }^{\left(  -  \right)} } & {\Delta _{{\bf{k}}, - s, - \sigma }^{\left(  +  \right)} }  \\
   {\Delta _{{\bf{k}},s,\sigma }^{\left(  +  \right)*} } & {\Delta _{{\bf{k}}, - s, - \sigma }^{\left(  -  \right)*} } & { - \xi _{\bf{k}} } & 0  \\
   {\Delta _{{\bf{k}},s,\sigma }^{\left(  -  \right)*} } & {\Delta _{{\bf{k}}, - s, - \sigma }^{\left(  +  \right)*} } & 0 & { - \xi _{\bf{k}} }  \\
\end{array}} \right)\left( {\begin{array}{*{20}c}
   {f_{{\bf{k}},s,\sigma } }  \\
   {f_{{\bf{k}}, - s, - \sigma } }  \\
   {f_{ - {\bf{k}},s,\sigma }^\dag  }  \\
   {f_{ - {\bf{k}}, - s, - \sigma }^\dag  }  \\
\end{array}} \right),
\eea
The mean fields are defined by
\be
\Delta _{{\bf k},s,\sigma }^{\left(  +  \right)}  = 
 + \frac{{2\pi i}}{{m\Omega }}s\sigma \sum\limits_{{\bf{k}} ( \ne {\bf{k'}} )} 
{\frac{{k_x k'_y  - k_y k'_x }}{{\left( {{\bf{k}} - {\bf{k}}'} \right)^2 }}
\left\langle {f_{ - {\bf k}',s,\sigma } f_{{\bf k}',s,\sigma } } \right\rangle },
\label{eq_mf_plus}
\ee
\be
\Delta _{{\bf k},s,\sigma }^{\left(  -  \right)}  =  
- \frac{{2\pi i}}{{m\Omega }}s\sigma \sum\limits_{{\bf{k}} ( \ne {\bf{k'}} )} 
{\frac{{k_x k'_y  - k_y k'_x }}{{\left( {{\bf{k}} - {\bf{k}}'} \right)^2 }}
\left\langle {f_{ - {\bf k}', - s, - \sigma } f_{{\bf k}',s,\sigma } } \right\rangle }.
\label{eq_mf_minus}
\ee
The mean field (\ref{eq_mf_plus}) is associated with the Cooper pairing
between holes with the same spin and the same half-Skyrmion index.
On the other hand, the mean field (\ref{eq_mf_minus}) is associated with 
the Cooper pairing between holes with the opposite spin 
and the opposite half-Skyrmion index.

If we consider the interaction (\ref{eq_interaction}) only,
then the pairing states described by $\Delta_{{\bf k},s,\sigma}^{(+)}$
and $\Delta_{{\bf k},s,\sigma}^{(-)}$ are degenerate energetically.
However, if we include the (anti-)half-Skyrmion-(anti-)half-Skyrmion 
interaction and the half-Skyrmion-anti-half-Skyrmion interaction,
the Cooper pairing between holes with the opposite
spin states and the opposite half-Skyrmion indices is favorable.
Because the interaction between (anti-)half-Skyrmions is repulsive
and the interaction between half-Skyrmions and anti-half-Skyrmions 
is attractive.
So we may focus on the pairing correlations described by 
$\Delta _{{\bf k},s,\sigma }^{\left(  -  \right)}$.

At zero temperature, the BCS gap equation is
\[
\Delta _{{\bf{k}},s,\sigma }^{\left(  -  \right)}  =  
- \frac{{\pi i}}{{m\Omega }}s\sigma \sum\limits_{{\bf{k}} ( \ne {\bf{k'}} )} 
{\frac{{k_x k'_y  - k_y k'_x }}{{\left( {{\bf{k}} - {\bf{k}}'} \right)^2 }}
\frac{{\Delta _{{\bf{k}}',s,\sigma }^{\left(  -  \right)} }}{{E_{{\bf{k}}'} }}},
\]
where $E_{\bf k}$ is the quasiparticle excitation energy.
This gap equation is divided into two equations according to
the relative sign between $s$ and $\sigma$.
For $\Delta_{{\bf k},s,s}^{(-)} \equiv \Delta_{{\bf k},+}^{(-)}$,
the gap equation is
\be
\Delta _{\bf{k},+}^{\left( - \right)}  =  
- \frac{{\pi i}}{{m\Omega }}\sum\limits_{{\bf{k}} ( \ne {\bf{k'}} )} 
{\frac{{k_x k'_y  - k_y k'_x }}{{\left( {{\bf{k}} - {\bf{k}}'} \right)^2 }}
\frac{{\Delta _{{\bf{k}}',+}^{\left( - \right)} }}{{E_{{\bf{k}}'} }}}.
\label{eq_gap_p}
\ee
For $\Delta_{{\bf k},s,-s}^{(-)} \equiv \Delta_{{\bf k},-}^{(-)}$,
the gap equation is
\be
\Delta _{\bf{k},-}^{\left( - \right)}  =  + \frac{{\pi i}}{{m\Omega }}
\sum\limits_{{\bf{k}} ( \ne {\bf{k'}} )} {\frac{{k_x k'_y  - k_y k'_x }}
{{\left( {{\bf{k}} - {\bf{k}}'} \right)^2 }}
\frac{{\Delta _{{\bf{k}}',-}^{\left( - \right)} }}{{E_{{\bf{k}}'} }}}.
\label{eq_gap_m}
\ee
Here we consider the case in which 
both of $\Delta_{{\bf k},+}^{(-)}$ and $\Delta_{{\bf k},-}^{(-)}$
describe the same pairing symmetry.
Under this condition, we find
$|\Delta_{{\bf k},+}^{(-)}|=|\Delta_{{\bf k},-}^{(-)}|$,
and so $E_{\bf k}$ is given by
\be
E_{\bf{k}}  = \sqrt {\xi _{\bf k}^2  + \left| 
{\Delta _{\bf{k},+}^{\left( - \right)} } 
\right|^2 }
= \sqrt {\xi _{\bf k}^2  + \left| 
{\Delta _{\bf{k},-}^{\left( - \right)} } 
\right|^2 }.
\ee

A similar gap equation was analyzed in Ref.~\refcite{Greiter1992}
in the context of the fractional quantum Hall systems.
Following Ref.~\refcite{Greiter1992}, we introduce an ansatz,
\be
\Delta_{{\bf k},+}^{(-)} = \Delta_k \exp \left(-i \ell \theta_{\bf k} \right),
\ee
where $\ell$ is an integer and $\theta_{\bf k}=\tan^{-1} (k_y/k_x)$.
Substituting this expression into \eref{eq_gap_p},
the integration with respect to the angle $\theta_{{\bf k}'}$ 
is carried out analytically.
At this procedure, we find that there is no solution for $\ell=0$.
Therefore, there is no $s$-wave pairing state.
Furthermore, the gap equation has the solutions only for $\ell > 0$.
The gap $\Delta_k$ satisfies the following equation,
\be
\Delta _k  = \frac{1}{{2m}}\int_0^k {dk'} \frac{{k'\Delta _{k'} }}{{E_{k'} }}
\left( {\frac{{k'}}{k}} \right)^\ell   + \frac{1}{{2m}}\int_k^\infty  {dk'} 
\frac{{k'\Delta _{k'} }}{{E_{k'} }}\left( {\frac{k}{{k'}}} \right)^\ell.
\label{eq_gap}
\ee
From the asymptotic forms in $k \rightarrow \infty$ and $k \rightarrow 0$,
we assume the following form for $\Delta_k$,
\[
\Delta _k /\varepsilon _F  = \left\{ \begin{array}{l}
 \Delta \left( {k/k_F } \right)^\ell  ,\left( {k < k_F } \right), \\ 
 \Delta \left( {k_F /k} \right)^\ell  ,\left( {k > k_F } \right), \\ 
 \end{array} \right.
\]
where $\epsilon_F$ is the Fermi energy of holes
and $k_F$ is the Fermi wave number.
The gap $\Delta$ is found numerically.
The gap $\Delta$ decreases by increasing $\ell$.
We find $\Delta=0.916$ for $\ell=1$
and $\Delta=0.406$ for $\ell=2$.

The gap equation (\ref{eq_gap_m}) is analyzed similarly.
However, because of the sign difference in the interaction
the solution has the following form,
\be
\Delta_{\bf k}^{(-)} = \Delta_k \exp \left(i \ell \theta_{\bf k} \right),
\ee
where $\Delta_k$ is the solution of \eref{eq_gap}.
As a result, there are two types of Cooper pairs
with opposite relative angular momentum.

For $p$-wave ($\ell=1$) gap symmetry,
the sum of $\Delta_{{\bf k},+}^{(-)}$ and
$\Delta_{{\bf k},-}^{(-)}$ leads to a $p_x$-wave gap
which is unstable in the bulk in the absence
of symmetry breaking associated with spatial anisotropy.
Since $s$-wave gap symmetry is ruled out as mentioned above,
the lowest energy sate is obtained for 
$d_{x^2-y^2}$-wave gap symmetry.

\begin{figure}
\centerline{\psfig{file=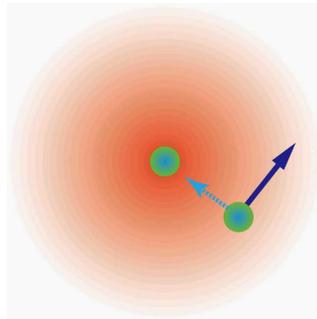,width=4.2cm}}
\caption{
Interaction between doped holes arising 
from a Berry phase effect associated with
the gauge flux created by a half-Skyrmion.
}
\label{fig_interaction}
\end{figure}

The pairing mechanism based on half-Skyrmions
is intuitively understood (\fref{fig_interaction}).
According to \eref{eq_density_flux}, a half-Skrymion, or
a gauge flux, is induced around a hole.
If another hole passes the gauge flux region at the Fermi velocity,
a magnetic Lorentz-force-like interaction acts on that hole
according to the interaction represented by \eref{eq_minc}.
A similar pairing interaction is discussed
at half-filled Landau levels.\cite{Greiter1992,Morinari2000}
In that system the gauge field is 
the Chern-Simons gauge field \cite{Zhang1992} whose gauge fluxes
cancel the external magnetic field fluxes at the mean field level.
Gauge field fluctuations give rise to a paring interaction between 
flux attached fermions.

\section{Pseudogap}
\label{sec_pseudogap}
One of the most intriguing phenomena observed in high-temperature
superconductors is the so-called pseudogap which is observed
in various physical quantities.\cite{Timusk1999}
Here we focus on the pseudogap behavior observed in ARPES.
If the hole concentration is lower than the optimum hole concentration
at which the transition temperature is the maximum,
the Fermi surface is not a conventional Fermi surface
expected from the band theory.
Instead, a truncated, an ark-like Fermi surface is observed 
in ARPES.\cite{Norman1998}
In order to explain this Fermi arc,
a standard approach is to consider a coupling with some boson modes, 
such as spin fluctuations or gauge field fluctuations associated with 
phase fluctuations of a mean field, expecting self-energy effects 
in the single body quasiparticle Green's function. 
However, it is not obvious that such a conventional analysis leads to 
qualitatively different physics.

The half-Skyrmion theory provides a completely different approach.
To describe the doped hole dynamics
which is associated with the spectral function observed by ARPES,
we need to include the fact that each doped hole carries 
a half-Skyrmion.
For that purpose, the direct way is to take 
a doped hole and the half-Skyrmion created by the hole
as a composite particle so that
the dynamics of the half-Skyrmion is included
in the doped hole dynamics.\cite{Morinari2008fa}

The Hamiltonian describing the hole hopping is
\be
H_{\mathrm{t}} = -\sum_{\langle i,j \rangle} \sum_{\sigma}
t_{ij} c_{i\sigma}^{\dagger} c_{j\sigma} + h.c.,
\ee
where $t_{ij}=t$ for the nearest neighbor sites, $t_{ij}=t_1$
for the next nearest neighbor sites, and $t_{ij}=t_2$
for the third nearest neighbor sites.
The parameters $t_1/t$ and $t_2/t$ are chosen so that 
the Fermi surface in the Fermi liquid phase is reproduced.\cite{Tohyama2000}

Now let us include the half-Skyrmion dynamics.
The half-Skyrmion dispersion is described by the Hamiltonian,
\[
H_{\mathrm{hs}}  = \sum\limits_{{\bf{k}} \in {\mathrm RBZ}} 
\sum_{\sigma}
{
\left( {\begin{array}{*{20}c}
   {c_{{\mathrm{e}}{\bf{k}}\sigma }^\dag  } & {c_{{\mathrm{o}}{\bf{k}}\sigma }^\dag  }  \\
\end{array}} \right)\left( {\begin{array}{*{20}c}
   0 & {\kappa _{\bf{k}} }  \\
   {\kappa _{\bf{k}}^* } & 0  \\
\end{array}} \right)\left( {\begin{array}{*{20}c}
   {c_{{\mathrm{e}}{\bf{k}}\sigma } }  \\
   {c_{{\mathrm{o}}{\bf{k}}\sigma } }  \\
\end{array}} \right)},
\]
where the ${\bf k}$-summation is taken over the reduced 
Brillouin zone, $|k_x \pm k_y| \le \pi$, 
$c_{{\mathrm{e}}{\bf k}\sigma} 
= (c_{{\bf k}\sigma}+c_{{\bf k}+{\bf Q},\sigma})/\sqrt{2}$
and 
$c_{{\mathrm{o}}{\bf k}\sigma} 
= (c_{{\bf k}\sigma}-c_{{\bf k}+{\bf Q},\sigma})/\sqrt{2}$,
and
\[
\kappa _{\bf{k}}  = -v\left[ {\left( {\cos k_x  + \cos k_y } \right) 
+ i\left( {\cos k_x  - \cos k_y } \right)} \right].
\]
The half-Skyrmion dispersion is given by $\pm |\kappa_{\bf k}|$.
This dispersion corresponds to \eref{eq_hs_disp} 
with $E_0=0$ and $c_{\mathrm{sw}} = v$.
The half-Skyrmion creation energy $E_0$ is zero because
it vanishes in the absence of the antiferromagnetic
long-range order.\cite{Auerbach1991}
The spin-wave velocity for the doped system is denoted by $v$
which is different from $c_{\mathrm{sw}}$ in the 
undoped system.
Here we use the same creation operators and 
the annihilation operators
for doped holes and half-Skyrmions.
Because a doped hole and the half-Skyrmion carried by the hole 
is taken as a composite particle.
Note that it is not necessary to distinguish between half-Skyrmions and 
anti-half-Skyrmions.
Because their excitation spectra are the same.

The dispersion energy of the composite particle 
is calculated from $H = H_{\mathrm{t}} + H_{\mathrm{hs}}$ as,
\[
E_{\bf{k}}^{\left(  \pm  \right)}  = \varepsilon _{\bf{k}}^{\left(  +  \right)}  
\pm \left| {\kappa _{\bf{k}}  + \varepsilon _{\bf{k}}^{\left(  -  \right)} } 
\right|,
\]
with 
$\varepsilon _{\bf{k}}^{\left(  \pm  \right)}  
= \left( {\varepsilon _{\bf{k}}  
\pm \varepsilon _{{\bf{k}} + {\bf{Q}}} } \right)/2$
and
$\varepsilon _{\bf{k}}  =  - 2t\left( {\cos k_x  + \cos k_y } \right) 
- 4t_1 \cos k_x \cos k_y  - 2t_2 \left( {\cos 2k_x  + \cos 2k_y } \right)$.
The spectral function is calculated following a standard 
procedure.\cite{AGD}
The imaginary time Green's function for up-spin is defined as
\[
G_{{\bf{k}} \uparrow } \left( \tau  \right) =  
- \left\langle {T_\tau  c_{{\bf{k}} \uparrow } 
\left( \tau  \right)c_{{\bf{k}} \uparrow }^\dag  
\left( 0 \right)} \right\rangle,
\]
where $c_{{\bf k} \uparrow}(\tau) 
= {\rm e}^{\tau (H - \mu N )} c_{{\bf k} \uparrow}
{\rm e}^{-\tau (H - \mu N )}$ 
with $\mu$ the chemical potential and $N$ the number operator
and $T_{\tau}$ is the
imaginary time ($\tau$) ordering operator.

The Matsubara Green's function is obtained by
\[
G_{{\bf{k}} \uparrow } \left( {i\omega _n } \right) 
= \int_0^{(k_B T)^{-1}}  {d\tau } e^{i\omega _n \tau } 
G_{{\bf{k}} \uparrow } \left( \tau  \right),
\]
where $\omega_n=\pi (2 n+1) k_B T$ ($n = 0, \pm 1, \pm 2, ...$)
is the fermion Matsubara frequensy.
By the analytic continuation, $i \omega_n \rightarrow \omega + i\delta$,
with $\delta$ an infinitesimal positive number,
the retarded Green's function is obtained.
Thus, the spectral function is
\bea
A_{\bf{k}} \left( \omega  \right) 
&=&  - \frac{1}{\pi }{\mathop{\rm Im}\nolimits} G_{{\bf{k}} \uparrow } 
\left( {\omega  + i\delta } \right) \nonumber \\
&=& 
\frac{1}{2}\left( {1 + \zeta _{\bf{k}} } \right)\delta 
\left( {\omega  - E_{\bf{k}}^{\left(  +  \right)} } + \mu \right) 
+ \frac{1}{2}\left( {1 - \zeta _{\bf{k}} } \right)\delta 
\left( {\omega  - E_{\bf{k}}^{\left(  -  \right)} } + \mu \right),
\label{eq_A}
\eea
where,
\[
\zeta _{\bf{k}}  = \frac{{ - \left( {2t + v} \right)
\left( {\cos k_x  + \cos k_y } \right)}}{{\sqrt {\left( {2t + v} \right)^2 
\left( {\cos k_x  + \cos k_y } \right)^2  + v^2 
\left( {\cos k_x  - \cos k_y } \right)^2 } }}.
\]

\begin{figure}[ht]%
\begin{center}
  \parbox{2.1in}{\epsfig{figure=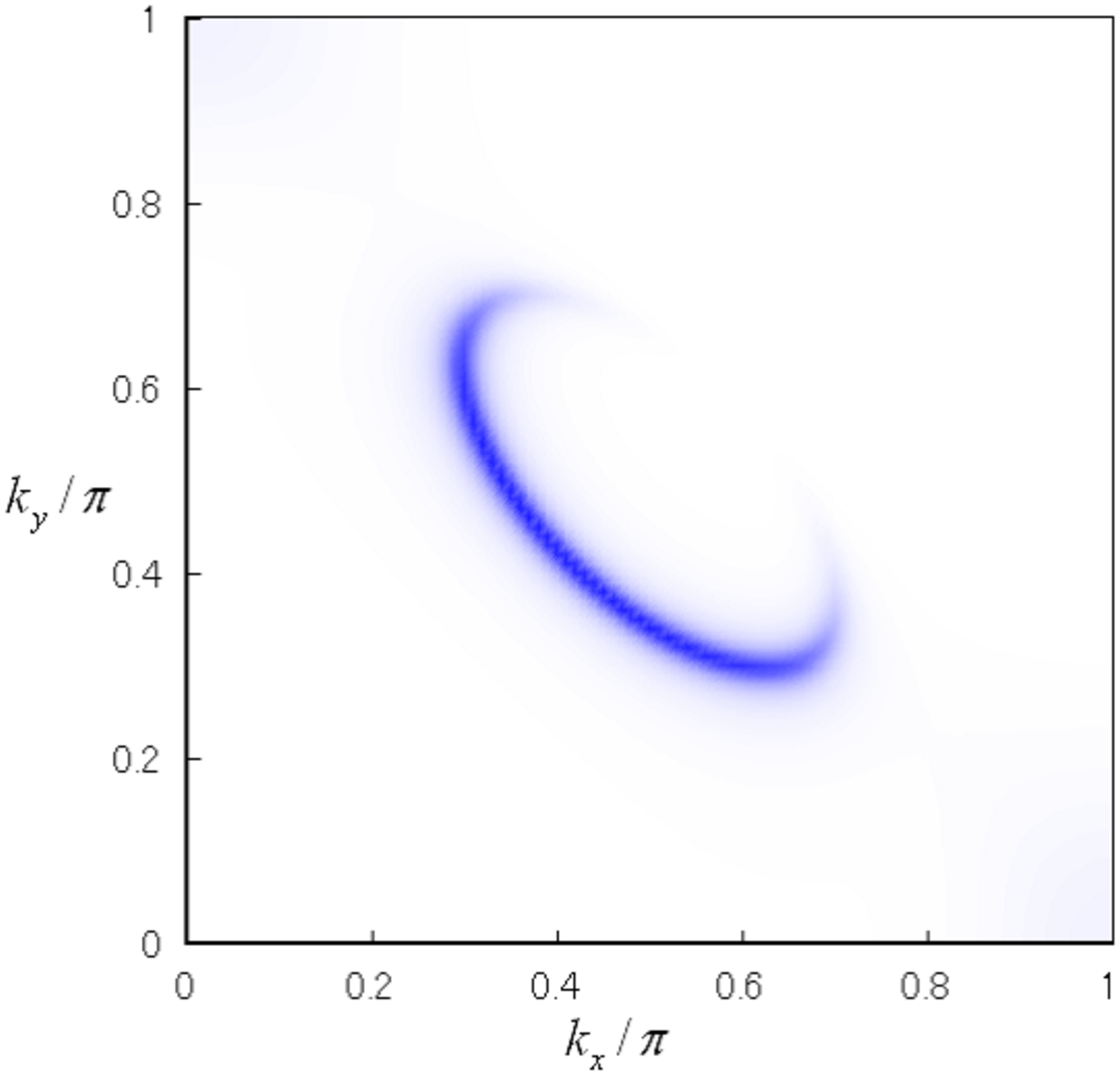,width=2in}\figsubcap{a}}
  \hspace*{4pt}
  \parbox{2.1in}{\epsfig{figure=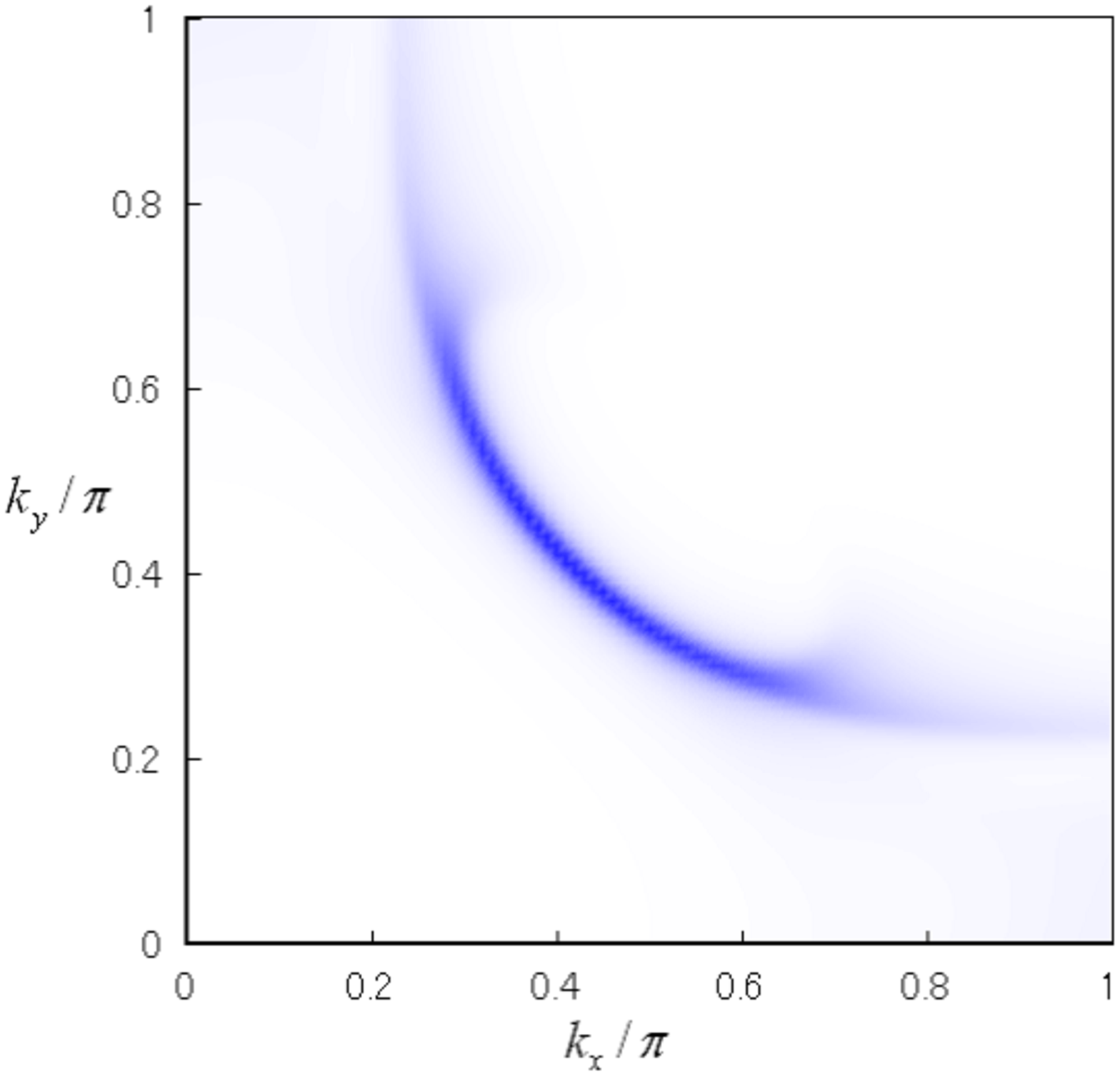,width=2in}\figsubcap{b}}
  \caption{
  (a) The intensity plot of the spectral function $A_{\bf k}(\omega=0)$
  in a quadrant of the Brillouin zone. 
  The doping concentration is $x=0.10$.
  The other parameters are $t_1/t=-0.25$,
  $t_2/t=0.10$, and $v/t=1.0$.
  (b) The intensity plot of the spectral function $A_{\bf k}(\omega=0)$
	with including the effect of the short-range antiferromagnetic correlation.
	The antiferromagnetic correlation length is taken as $\xi_{AF}=10$.
}  
\label{fig_fermi_arc}
\end{center}
\end{figure}

In numerical computations, 
the parameter $\delta$ above is taken to be finite
so that the $\delta$-function in the right hand side of \eref{eq_A}
is replaced by the Lorentz function.
\Fref{fig_fermi_arc}(a) shows $A_{\bf k}(\omega=0)$ 
with $\delta/t=0.10$.
The resulting Fermi surface is arc-like because the factors
$1\pm \zeta_{\bf k}$ suppress the intensity in part of
the Brillouin zone.
This arc-like Fermi surface is consistent with
the ARPES results.

However, there is some deviation from the experiment 
around the ends of the arc.\cite{Norman2007}
Basically the Fermi arc follows the underlying Fermi
surface which appears at high-temperature above the characteristic 
pseudogap temperature, $T^*$.
The deviation is suppressed by including the effect
of the short-range antiferromagnetic correlation.
\cite{Harrison2007,Morinari2008sraf}
The effect is included by taking average over the wave vector
change in the (incommensurate) antiferromagnetic correlation.
The calculation is similar to that described 
in Ref.~\refcite{Harrison2007}.
\Fref{fig_fermi_arc}(b) shows how the spectral intensity
is modified at the antiferromagnetic correlation length $\xi_{AF}=10$.

Now we comment on the similarity to 
the d-density wave theory.\cite{Chakravarty2001}
The calculation for the spectral function $A_{\bf k}(\omega)$
is almost identical to that in the d-density wave theory.\cite{Chakravarty2003}
However, the physical interpretation of $v$ is different.
In d-density wave theory, $v$ is associated 
with d-density wave long-range order.
But here $v$ is associated with the half-Skyrmions dynamics.
Although the origin is different, the half-Skyrmion theory
and the d-density wave theory
may share some results about
the pseudogap phenomenon because the Hamiltonian is almost the same.

\section{Summary}
\label{sec_summary}
In this chapter, we have reviewed the half-Skyrmion theory
for high-temperature superconductivity.
We have mainly focused on four topics.
First, we have discussed the single hole doped system.
The correlation of forming a singlet state between a doped hole spin
and a copper spin has been investigated including the fact 
that the hole wave function extends over a space,
and affected by the other copper site spins.
A half-Skyrmion spin texture created by
a doped hole has been described.
It is shown that the half-Skyrmion excitation spectrum 
is in good agreement with the ARPES results in the undoped system.

The most important aspect of the half-Skyrmion
is that the doped hole carries a topological charge
which is represented by a gauge flux in the CP$^1$ representation.
This property is in stark contrast to conventional spin polaron pictures.
Because in that case interaction clouds arising from spin correlations
are not characterized by a topological charge or a gauge flux.

Second, we have considered multi-half-Skymion configurations.
In multi-half-Skyrmion configurations,
antiferromagnetic long-range order is suppressed
around the doping concentration, $x = 0.10$.
The topological property of half-Skymions 
is kept for $x<0.30$ as shown by numerical simulations.
After suppression of antiferromagnetic long-range order,
the spin correlation becomes incommensurate.
Numerical simulations suggest that 
the origin of the incommensurate spin correlation
is associated with stripe configurations of half-Skyrmions
and anti-half-Skyrmions.

Third, we have discussed a mechanism of $d_{x^2-y^2}$-wave 
superconductivity.
The gauge flux created by half-Skyrmions induces
the interaction between doped holes.
The interaction leads to a $d_{x^2-y^2}$-wave superconducting state
of doped holes.
The origin of the attractive interaction is a Lorentz force
acting on a hole moving in a gauge flux created by 
another hole.

Finally, we have discussed a pseudogap phenomenon.
We have considered a composite particle of a hole and half-Skyrmion.
The pseudogap is associated with the excitation
spectrum of the composite particle.

Although several aspects of the half-Skyrmion theory
for high-temperature superconductivity are described 
in this review,
we need further studies to establish the theory.
In particular it is necessary to show the half-Skyrmion
formation in the single hole doped system in a more
convincing way to provide a sound starting point.

\section*{Acknowledgements}
I would like to thank T. Tohyama and G. Baskaran for helpful discussions.
This work was supported by the Grant-in-Aid for Scientific Research
from the Ministry of Education, Culture, Sports, Science and Technology (MEXT) 
of Japan, 
the Global COE Program "The Next Generation of Physics, 
Spun from Universality and Emergence," 
and Yukawa International Program for Quark-Hadron Sciences at YITP.
The numerical calculations were carried out in part 
on Altix3700 BX2 at YITP in Kyoto University.

\bibliographystyle{ws-rv-van}
\bibliography{../../../references/tm_library2}
\printindex
\end{document}